\documentclass[letterpaper, 10 pt, conference]{ieeeconf}  

\IEEEoverridecommandlockouts                              
\usepackage{amsmath,amsfonts,amssymb,graphicx,hyperref}
\usepackage[utf8]{inputenc}
\usepackage{bbold}
\usepackage{color,tikz}

\title{\LARGE \bf
Efficient learning of arbitrary single-copy quantum states
}

\author{Shibdas Roy$^{*,1,2,3}$, Filippo Caruso$^3$, Srushti Patil$^{4,1}$ and Anumita Mukhopadhyay$^{1,2}$
\thanks{This work was financially supported by the European Union's Horizon 2020 research and innovation programme under FET-OPEN Grant Agreement No.~828946 (PATHOS). SR thanks Sreetama Das for useful feedback.}
\thanks{$^1$Centre for Quantum Engineering, Research and Education (CQuERE), TCG Centres for Research and Education in Science and Technology (TCG CREST), Kolkata 700091, India; $^2$Academy of Scientific and Innovative Research (AcSIR), Ghaziabad 201002, India; $^3$Department of Physics and Astronomy, University of Florence, 50019 Sesto Fiorentino, Italy; $^4$Novo Nordisk Foundation Quantum Computing Programme, Niels Bohr Institute, University of Copenhagen, 2100 København Ø, Denmark.
        $^{*}${\tt\small Email: roy.shibdas@gmail.com}}%
}

\begin{document}

\maketitle
\thispagestyle{empty}
\pagestyle{empty}

\begin{abstract}

Quantum state tomography is the problem of estimating a given quantum state. Usually, it is required to run the quantum experiment - state preparation, state evolution, measurement - several times to be able to estimate the output quantum state of the experiment, because an exponentially high number of copies of the state is required. In this work, we present an efficient algorithm to estimate with a small but non-zero probability of error the output state of the experiment using a single copy of the state, without knowing the evolution dynamics of the state. It also does not destroy the original state, which can be recovered easily for any further quantum processing. As an example, it is usually required to repeat a quantum image processing experiment many times, since many copies of the state of the output image are needed to extract the information from all its pixels. The information from $\mathcal{N}$ pixels of the image can be inferred from a single run of the image processing experiment in our algorithm, to efficiently estimate the density matrix of the image state.

\end{abstract}

\section{Introduction}
A typical quantum experiment comprises state preparation, state evolution and measurement, before post-processing the measurement data \cite{NC}. The experiment needs to be repeated several times to be able to post-process the measurement data to estimate the evolved state. The higher number of times the experiment is repeated, the more accurate will be the estimation of the state. This is indeed a big bottleneck in performing many tasks using quantum computers. The key reason for the problem is that the measurement of a quantum state collapses the state. To deduce the output quantum state of an experiment, one needs to repeat the experiment and perform several measurements, even for a small number of qubits. This is because multiple copies of the state are required to infer the state, and since no-cloning theorem mandates that an arbitrary quantum state cannot be copied, the experiment needs to be repeated many times, exponentially high in the number of qubits.

Quantum amplitude estimation is an important subroutine in many quantum algorithms, and involves estimating the probability amplitudes of a given quantum state \cite{BHMT,GGZW,PRFL,TNSY,SURTOY}. It is possible to estimate $\mathcal{N}$ probability amplitudes in a quantum state by repeating the experiment $\mathcal{N}-1$ times, and every run of the experiment achieves a quadratic speedup with quantum amplitude estimation over classical estimation strategies, by using quantum amplitude amplification, based on Grover's search \cite{BHMT}. This algorithm uses phase estimation subroutine \cite{NC} to perform quantum amplitude estimation and requires that we have the unitary determining the evolution dynamics of the state. As a consequence, despite a speedup in the estimation process, it is still required to run the entire experiment $\mathcal{N}-1$ times. Note that a general quantum state of $N$ qubits has $\mathcal{N}=2^N$ probability amplitudes, and so, the $\mathcal{N}-1$ number of times that the experiment has to be run increases exponentially with increasing $N$.

Quantum amplitude estimation can be performed only for a given pure quantum state. Quantum state tomography, by contrast, is the problem of estimating a general quantum state, which can be mixed as well \cite{DPS,PR,SA2,RS,GFF,LWXHF,CPFSG}. In practical scenarios, particularly in the current noisy intermediate-scale quantum (NISQ) era, quantum states are inevitably noisy. However, existing methods of quantum tomography require multiple copies of the state, and so, require the entire experiment to be repeated as many times. In this work, we present an efficient quantum state tomography algorithm to approximately estimate the density matrix of an output quantum state of an experiment with a single copy of the state, without knowing anything about the evolution dynamics of the state. Our algorithm requires only a single run of the experiment, after which the $\mathcal{N}\times\mathcal{N}$ density matrix of the state can be efficiently estimated in $O({\rm poly} N)$ steps, with a non-zero probability of error. Since we do not perform measurements on the original state, we do not destroy the state, that can be restored for any further processing.

\section{Problem}
Let us consider a quantum experiment that starts with an input state $|0\rangle$. The action of the experiment is to apply an arbitrary unitary $U$ to the input state, and perform measurements on the output state $U|0\rangle$. The measurement data is then post-processed classically to obtain the desired information. Our goal here is to estimate the probability amplitudes of the state $U|0\rangle$. Let us denote $|\psi\rangle = U|0\rangle = \sqrt{a}|\psi_1\rangle + \sqrt{1-a}|\psi_0\rangle$, where $a$ is the amplitude of the state $|\psi_1\rangle$ that we try to estimate. One can amplify the probability of obtaining the state $|\psi_1\rangle$ by applying the operator $Q = -US_0U^{-1}S_X$ \cite{BHMT,SURTOY}, where the operator $S_X$ conditionally changes the sign of the amplitude of $|\psi_1\rangle$, but does nothing to $|\psi_0\rangle$. The operator $S_0$ changes the sign of the amplitude of $|0\rangle$. The space spanned by the vectors $|\psi_1\rangle$ and $|\psi_0\rangle$ has an orthonormal basis containing two eigenvectors of $Q$ \cite{BHMT}:
\begin{equation}
|\psi_{\pm}\rangle = \frac{1}{\sqrt{2}}\left(\frac{1}{\sqrt{a}}|\psi_1\rangle \pm \frac{i}{\sqrt{1-a}}|\psi_0\rangle\right),
\end{equation}
and the corresponding eigenvalues are  $\omega_{\pm} = e^{\pm i2\theta_a}$, where $i=\sqrt{-1}$ is the imaginary unit, and $\sin^2(\theta_a) = a$, such that $0\leq\theta_a\leq\pi/2$. Then, $|\psi\rangle$ can be expressed as:
\begin{equation}
|\psi\rangle = U|0\rangle = \frac{-i}{\sqrt{2}}\left(e^{i\theta_a}|\psi_{+}\rangle-e^{-i\theta_a}|\psi_{-}\rangle\right).
\end{equation}

The amplitude estimation algorithm from Ref.~\cite{BHMT} is: (i) Initialize two registers to the state $|0^{\otimes\beta}\rangle|\psi\rangle$. (ii) Apply quantum Fourier transform to first register. (iii) Apply the operation: $|b\rangle|y\rangle \mapsto |b\rangle\left(Q^b|y\rangle\right)$, where $0\leq b<M=2^\beta$. (iv) Apply inverse quantum Fourier transform to first register. (v) Measure first register to obtain the outcome $|\phi\rangle$. (vi) Output the estimate of $a$ as $\tilde{a}=\sin^2\left(\pi\frac{\phi}{M}\right)$. This algorithm clearly requires performing phase estimation using controlled-$Q$ operations to obtain an accurate estimate of $a$. An alternative method without using phase estimation from Ref.~\cite{SURTOY} also achieves a quadratic quantum speedup asymptotically, but without requiring controlled-$Q$ operations. Hence, quantum amplitude estimation takes $O(\sqrt{\mathcal{N}}/\varepsilon)$ $=$ $O(2^{N/2}/\varepsilon)$ steps to additive accuracy $\varepsilon$, that is still exponential in $N$.

The approach to quantum amplitude estimation discussed above requires to have the unitary $U$ to be able to determine the operator $Q$, required for performing the phase estimation using controlled-$Q$ operations. However, in general, consider that Alice transmits a quantum state through a quantum channel to Bob. There may or may not be other quantum operations on the quantum state within the channel during the transmission. Bob needs to know the density matrix of the state, without knowing anything about the evolution dynamics of the channel that the transmitted quantum state went through. The channel could well be a unitary or a noisy channel that Bob has no idea about. All that Bob has at the receiving end is a quantum state, and he wants to know the elements of the density matrix of the state. Usually, Alice and Bob would need to repeat the entire process several times to be able to deduce just the diagonal elements of the density matrix. Our aim is to estimate all the elements of the density matrix of the output state, without repeating the process at all, or without knowing anything about the channel dynamics.

\section{Method}
Consider that the number of qubits in the output quantum state is $N$, so that $\mathcal{N}=2^N$. We denote the density matrix of the output state as $\rho$. If the input state is $\rho(0)$ and the Hamiltonian of the channel is $H$, then we get \cite{NC}:
$\dot{\rho}(t) = \frac{-i}{\hbar}[H,\rho]$, such that we can define a superoperator:
\begin{equation}
\rho(t) = e^{\frac{-i}{\hbar}[H,\rho]t}\left[\rho(0)\right] = e^{\frac{i}{\hbar} Ht}\rho(0)e^{-\frac{i}{\hbar} Ht},
\end{equation}
where $\hbar$ is the reduced Planck's constant, and $[H,\rho]=H\rho-\rho H$ is a commutator. Note that if we have $H=\rho$, the commutator is zero, and so, the superoperator, being identity, has an effect of keeping the state unchanged, i.e.~$e^{\frac{i}{\hbar}\rho}\rho e^{\frac{-i}{\hbar}\rho} = \rho$, where we dropped the time variable $t$, since for $H=\rho$, we have $\dot{\rho}(t)=0$. Thus, the unitary $V:=e^{\frac{i}{\hbar}\rho}$ is not the identity operator, but has a similar effect of keeping $\rho$ unchanged. The state $\rho$ itself can act as a Hamiltonian to yield the unitary $V$, which, in turn, has no effect on the state itself. This also follows from the Baker-Hausdorff lemma, with $G=J=\rho$, $\gamma=\frac{1}{\hbar}$ \cite{SNbook}:
\begin{equation}
e^{iG\gamma}Je^{-iG\gamma}=J+i\gamma[G,J]+\frac{i^2\gamma^2}{2!}[G,[G,J]]+\ldots
\end{equation}

\begin{figure*}[!t]
    \centering
    \includegraphics[width=0.87\linewidth]{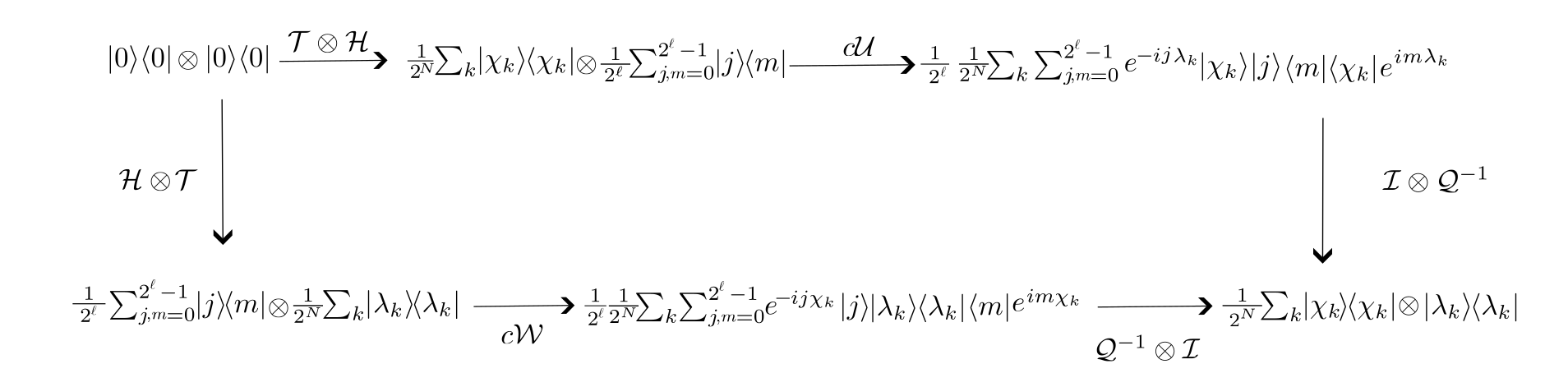}
    \caption{Two ways of evolving $|0\rangle|0\rangle$ to $\zeta$, taking $\ell=N$ and assuming perfect phase estimations, i.e.~$\tilde{\chi}_k=\chi_k$, $\tilde{\lambda}_k=\lambda_k$ $\forall k$.}
    \label{fig:qst}
\end{figure*}

Exponentiation of an unknown density matrix $\rho$ can be performed efficiently in $O({\rm poly} N)$ steps with $n$ copies of $\rho$ \cite{LMR,KSGS}. Given two registers $C$ and $D$ initialized to the unknown $\rho$ and a known $\sigma$, respectively, the below is applied repeatedly, to construct $e^{\frac{-i}{\hbar}\rho n\Delta t}\sigma e^{\frac{i}{\hbar}\rho n\Delta t}$ \cite{LMR}:

\small
\begin{equation}\label{eq:exprho_step}
{\rm Tr}_C\left[e^{\frac{-i}{\hbar}\mathcal{S}\Delta t}\left(\rho\otimes\sigma\right)e^{\frac{i}{\hbar}\mathcal{S}\Delta t}\right] = \sigma -\frac{i}{\hbar}\Delta t[\rho,\sigma] +O(\Delta t^2),
\end{equation}\normalsize
where $\mathcal{S}$ is the swap operator, which is sparse and so, $e^{\frac{-i}{\hbar}\mathcal{S}\Delta t}$ can be performed efficiently \cite{BACS,HHL}. The number of copies $n$ of $\rho$ required is of the order $n=O\left(\frac{t^2}{\hbar^2}\epsilon^{-1}\right)$, where $t=n\Delta t$, and $\epsilon$ is the desired accuracy. Thus, only $n$ copies are known to be required to construct $e^{\frac{-i}{\hbar}\rho t}$, where $n$ is independent of the dimension of $\rho$, and the circuit depth required is $O(Nt^2\hbar^{-2}\epsilon^{-1})$ \cite{KSGS}. Then, the unitary $\mathcal{V}:=e^{-i\rho\sqrt{\epsilon}}$ can be simulated efficiently with only $n=1$ copy of the matrix $\rho$, by taking $\Delta t=\hbar\sqrt{\epsilon}$, that is small enough for (\ref{eq:exprho_step}) to hold \cite{LMR}, for sufficiently small chosen value of $\epsilon$. This is because $\epsilon$ is given by the following equation (see Appendix H in Ref.~\cite{KSGS}):
\begin{equation}\label{eq:dme}
\epsilon = 1 - \cos^{2n}\left(\frac{t}{\hbar n}\right),
\end{equation}
which, for $n=1$, becomes $\epsilon = 1-\cos^2(t/\hbar) = \sin^2(t/\hbar)$,
that, in turn, for small $t/\hbar$, yields $\epsilon=t^2/\hbar^2 \Rightarrow t=\hbar\sqrt{\epsilon}$. Note that if the simulation error $\epsilon$ is an error in trace distance, then it determines the maximum probability of simulation error (see Ref.~\cite{KLLOY}), and so, it need not necessarily be $O(1/2^N)$ and can be $O(1/{\rm poly} N)$, as long as the overall probability of error of our algorithm is not larger than $1/3$.

Now, we take the known matrix $\sigma$ as a diagonal matrix, and upon applying the unitary $\mathcal{V}$, we get $\eta = e^{-i\rho\sqrt{\epsilon}}\sigma e^{i\rho\sqrt{\epsilon}}$. Thus, the matrix $\mathcal{V}$ diagonalizes the Hermitian matrix $\eta$ to $\mathcal{V}^{-1}\eta\mathcal{V}=\sigma$. Then, the eigenvalues of $\eta$ are just the diagonal elements of $\sigma$, and the eigenstates of $\eta$ are the columns of the matrix $\mathcal{V}$. We, therefore, want to find the eigenstates of the matrix $\eta$. We do this by performing exponentiation of the density matrix $\eta$, as we did of $\rho$ above, with $n=1$ copy of $\eta$, to construct the unitary $\mathcal{U}:=e^{-i\eta\sqrt{\epsilon}}$. If we, then, perform phase estimation of $\mathcal{U}$ using an initial state $\mathcal{I}/2^N\otimes|0^{\otimes\ell}\rangle\langle 0^{\otimes\ell}|$ in two registers $A$ and $B$, we would get at the output $\zeta := \frac{1}{2^N}\sum_k|\chi_k\rangle\langle\chi_k|\otimes|\tilde{\lambda}_k\rangle\langle\tilde{\lambda}_k|$ \cite{LMR}, where $|\chi_k\rangle$ are the eigenstates of $\eta$, $\tilde{\lambda}_k$ are estimates of the eigenvalues $\lambda_k$, and $\mathcal{I}/2^N=\frac{1}{2^N}\sum_k|\chi_k\rangle\langle\chi_k|$ is the maximally mixed state, that can be expressed in terms of any complete set of orthonormal eigenstates. Here, $\mathcal{I}$ is the identity operator. Let $\mathcal{T}:|0\rangle\langle 0| \to\mathcal{I}/2^N=\sum_l\mathcal{L}_l|0\rangle\langle 0|\mathcal{L}_l^\dagger$, so that $\mathcal{T}^{-1}:\mathcal{I}/2^N\to|0\rangle\langle 0|=\sum_l\mathcal{L}_l^\dagger(\mathcal{I}/2^N)\mathcal{L}_l$, where $\{\mathcal{L}_l\}$ are Kraus operators satisfying $\sum_l\mathcal{L}_l^\dagger\mathcal{L}_l=\mathcal{I}$. A single-qubit maximally-mixed state can be realized by creating a two-qubit maximally entangled state (using Hadamard and $CNOT$ gates), of which we take only one qubit, which is effectively in a maximally mixed state. A two-qubit maximally-mixed state is just two copies of the single-qubit maximally-mixed state, and similarly, an $N$-qubit maximally-mixed state is just $N$ copies of the single-qubit maximally-mixed state, i.e.~$(1/2^N)\sum_{k=0}^{2^N-1}|k\rangle\langle k|=\left[(1/2)\left(|0\rangle\langle 0|+|1\rangle\langle 1|\right)\right]^{\otimes N}$. We denote the phase estimation operator that yielded $\zeta$ from $|0\rangle|0\rangle$ as $\mathcal{M}:=\left(\mathcal{I}\otimes\mathcal{Q}^{-1}\right)c\mathcal{U}\left(\mathcal{T}\otimes\mathcal{H}\right)$, where $c\mathcal{U}$ is controlled-$\mathcal{U}$ operation, $\mathcal{H}$ is Hadamard gate of right dimension, used to create equal superposition, and $\mathcal{Q}$ is quantum Fourier transform (QFT). The controlled-$\mathcal{U}$ operation can be performed simply by replacing the swap operator with a conditional swap in (\ref{eq:exprho_step}) \cite{LMR}. That is, we act the conditional swap operator $|\tau\rangle\langle\tau|\otimes e^{-i\mathcal{S}\frac{t}{\hbar}\tau/T}$ on $|\tau\rangle\langle\tau|\otimes\varrho\otimes\eta$ for say, $\varrho=|\varpi\rangle\langle\varpi|$, and then take partial trace over $\eta$, to yield the conditional Hamiltonian evolution $|\tau\rangle\langle\tau|\otimes e^{-i\eta\frac{t}{\hbar}\tau/T}$, as required for \emph{improved} phase estimation to be discussed shortly. Note that the eigenvalues of $\mathcal{S}$ are $\pm 1$, so that $W^{1/T}=W^{2^1/T}=W^{2^2/T}=W^{2^3/T}\ldots$, where $W=e^{-i\mathcal{S}t/\hbar}$, if we use say $t=2\pi$.

Now, we have $\mathcal{I}/2^N=\frac{1}{2^N}\sum_k|\lambda_k\rangle\langle\lambda_k|$ for orthonormal $N$-qubit eigenvectors $|\lambda_k\rangle$. Let us denote with $c\mathcal{W}$, controlled-$\mathcal{W}$ operation, where $\mathcal{W}:=\sum_k e^{-i\chi_k}|\lambda_k\rangle\langle\lambda_k|$. Then, the effect of $\mathcal{K}:=\left(\mathcal{Q}^{-1}\otimes\mathcal{I}\right)c\mathcal{W}\left(\mathcal{H}\otimes\mathcal{T}\right)$ is equivalent to that of $\mathcal{M}$ on an initial state $|0\rangle|0\rangle$ with $\ell=N$ to yield $\zeta$. Please see Figure \ref{fig:qst}. So, $\mathcal{M}=\mathcal{K}$ implies $c\mathcal{W}=\left(\mathcal{Q}\otimes\mathcal{Q}^{-1}\right)c\mathcal{U}\left(\mathcal{T}\mathcal{H}\otimes\mathcal{H}\mathcal{T}^{-1}\right)$. We perform phase estimation of the eigenstates $|\lambda_k\rangle$ of $\mathcal{W}$, that we know from $\sigma$, to get the estimates $\tilde{\chi}_k$ of the phases $\chi_k$ of $\mathcal{W}$ for each $k$. We do so by applying the operator $\mathcal{R}:=\left(\mathcal{Q}^{-1}\otimes\mathcal{I}\right)c\mathcal{W}\left(\mathcal{H}\otimes\mathcal{I}\right)=\left(\mathcal{I}\otimes\mathcal{Q}^{-1}\right)c\mathcal{U}\left(\mathcal{T}\otimes\mathcal{H}\mathcal{T}^{-1}\right)$ to take $|0^{\otimes\ell}\rangle|\lambda_k\rangle$ to $|\tilde{\chi}_k\rangle|\lambda_k\rangle$, and then measuring the first register. These $\tilde{\chi}_k$ are columns of the estimate $\tilde{\mathcal{V}}$ of $\mathcal{V}$.

Ref.~\cite{LMR} showed that $n=O(1/\epsilon^3)$ copies of the state would be required to exponentiate it and perform phase estimation on the unitary, with the improved phase estimation technique from Ref.~\cite{HHL}, which is as follows. We start with an initial state $|\Lambda_0\rangle|u_j\rangle$, where $|u_j\rangle$ is the $j$-th eigenstate of the density matrix $\Gamma$, and $|\Lambda_0\rangle:=\sqrt{\frac{2}{T}}\sum_{\tau=0}^{T-1}\sin\frac{\pi(\tau+\frac{1}{2})}{T}|\tau\rangle$ for some large $T=2^\ell$. The initial state $|\Lambda_0\rangle$ can be prepared upto some error $\epsilon_\Lambda$ in time ${\rm poly}\log(T/\epsilon_\Lambda)$ (see Section A of Supplementary material of Ref.~\cite{HHL}). We apply the conditional Hamiltonian evolution $\sum_{\tau=0}^{T-1}|\tau\rangle\langle\tau|\otimes e^{i\Gamma\tau t/T}$ on the initial state in both registers, and then apply quantum Fourier transform (QFT) $\mathcal{Q}$ on the first register to obtain the state $\sum_{p=0}^{T-1}\alpha_{p|j}|p\rangle|u_j\rangle$. Defining the estimate $\tilde{r}_p$ of the $p$-th eigenvalue $r_p$ of $\Gamma$ as $\tilde{r}_p:=\frac{2\pi p}{t}$, we can relabel the Fourier basis states $|p\rangle$ to obtain $\sum_{p=0}^{T-1}\alpha_{p|j}|\tilde{r}_p\rangle|u_j\rangle$. If the phase estimation is perfect, we have $\alpha_{p|j}=1$ if $\tilde{r}_p=r_j$, and $0$ otherwise. So, we obtain the state $|\tilde{r}_j\rangle|u_j\rangle$, from which we get the estimate of $r_j$ upon measuring the first register. Thus, this improved method requires $t=O(1/\epsilon)$. Since the exponentiation of $\Gamma$ requires $n=O(t^2/\epsilon)$ copies of $\Gamma$, the overall phase estimation process requires $n=O(1/\epsilon^3)$ copies of $\Gamma$. However, replacing $t \to \nu t$ in (\ref{eq:dme}) yields $\epsilon = 1 - \cos^{2n}\left(\frac{\nu t}{\hbar n}\right)$, that for $n = 1$ becomes $\epsilon = \nu^2t^2/\hbar^2$. The improved estimation precision error $\epsilon$ satisfies $t=O(1/\epsilon)=2\pi^2/\epsilon$ (see Supplementary Material of Ref.~\cite{HHL}). Introducing the constant factor $\nu = \frac{\hbar\epsilon\sqrt{\epsilon}}{2 \pi^2}$, and taking $\nu t = \nu\Delta t$, we need just $n=1$ copy of $\Gamma$ with this method (e.g.~the unitary $\mathcal{V}$ is $e^{-i\rho\nu t/\hbar}=e^{-i\rho\sqrt{\epsilon}}$). 

Also, note that the error $\epsilon$ here, being twice the error in trace distance, that determines the maximum probability of estimation error (again see Supplementary Material of Ref.~\cite{HHL}), can be $O(1/{\rm poly}N)$ and not necessarily $O(1/2^N)$ (but using $T=2^N$), if we do not demand that the eigenvalues of a full-rank density matrix $\Gamma$ must all be resolved distinctly. Thus, $\epsilon$ depicts both the error in simulating the exponentiation of the density matrix $\Gamma$, and the (improved) phase estimation error in determining the eigenvalues of $\Gamma$, and is determined in both cases by an error in trace distance. Hence, $\epsilon$ determines the maximum probability of simulation or estimation error, that can scale as $O(1/{\rm poly}N)$ and not necessarily $O(1/2^N)$, as long as the overall error probability of our algorithm is less than or equal to $1/3$.

Let us denote the conditional Hamiltonian evolution discussed above as $c\mathcal{U}^\prime$ for the unitary $\mathcal{U}$, where the control state is inherently implied by the case under consideration. Then, the phase estimation operator $\mathcal{M}$ is modified to $\mathcal{M}^\prime=\mathcal{F}\left(\mathcal{T}\otimes\mathcal{G}\right)$, that takes the state $|0\rangle\langle 0|\otimes|0\rangle\langle 0|$ to $\zeta$, where $\mathcal{F}=\left(\mathcal{I}\otimes\mathcal{Q}\right)c\mathcal{U}^\prime$. Here, we have $\mathcal{G}|0\rangle=|\Lambda_0\rangle$. Similarly, the operator $\mathcal{K}$ modifies to $\mathcal{K}^\prime=\left(\mathcal{Q}\otimes\mathcal{I}\right)c\mathcal{W}^\prime\left(\mathcal{G}\otimes\mathcal{T}\right)$. Thus, $\mathcal{M}^\prime=\mathcal{K}^\prime$ implies $c\mathcal{W}^\prime=\left(\mathcal{Q}^{-1}\otimes\mathcal{Q}\right)c\mathcal{U}^\prime\left(\mathcal{T}\mathcal{G}^{-1}\otimes\mathcal{G}\mathcal{T}^{-1}\right)$. Accordingly, the operator $\mathcal{R}$ modifies to $\mathcal{R}^\prime=\left(\mathcal{Q}\otimes\mathcal{I}\right)c\mathcal{W}^\prime=\left(\mathcal{I}\otimes\mathcal{Q}\right)c\mathcal{U}^\prime\left(\mathcal{TG}^{-1}\otimes\mathcal{G}\mathcal{T}^{-1}\right)$, that takes the state $|\Lambda_0\rangle|\lambda_k\rangle$ to the state $|\tilde{\chi}_k\rangle|\lambda_k\rangle$ for each $k$.

Once the estimate $\tilde{\mathcal{V}}$ of the unitary $\mathcal{V}=e^{-i\rho\sqrt{\epsilon}}$ is obtained, one is left with simply finding out its eigenvalues and eigenvectors to determine the estimate $\tilde{\rho}$ of $\rho$. However, it may not, in general, be computationally efficient to find the $2^N$ numbers of eigenvalues $e^{-i\tilde{\kappa}_q\sqrt{\epsilon}}$ and eigenvectors $|\tilde{\xi}_q\rangle$ of $\tilde{\mathcal{V}}$, to, in turn, obtain the estimates $\tilde{\kappa}_q$ of the eigenvalues $\kappa_q$ and the estimates $|\tilde{\xi}_q\rangle$ of the eigenstates $|\xi_q\rangle$ of $\rho$. In order to obtain the estimate $\tilde{\rho}$ of $\rho$ from $\tilde{\mathcal{V}}$ efficiently, the maximally mixed state $\mathcal{I}/2^N$ is used as the input to perform phase estimation of $\tilde{\mathcal{V}}$, where now we have $\mathcal{I}/2^N=\frac{1}{2^N}\sum_q|\tilde{\xi}_q\rangle\langle\tilde{\xi}_q|$. Then, the output of the phase estimation will be: $\frac{1}{2^N}\sum_q|\tilde{\kappa}_q\rangle\langle\tilde{\kappa}_q|\otimes|\tilde{\xi}_q\rangle\langle\tilde{\xi}_q|$. Now, like in Ref.~\cite{HHL}, we add an ancilla qubit initialised in the state $\mathcal{I}/2=\frac{1}{2}(|0\rangle\langle 0|+|1\rangle\langle 1|)$, and rotate it conditioned on $|\tilde{\kappa}_q\rangle$ to get: $\frac{1}{2^N}\sum_q|\tilde{\kappa}_q\rangle\langle\tilde{\kappa}_q|\otimes|\tilde{\xi}_q\rangle\langle\tilde{\xi}_q|\otimes\left[(1-\tilde{\kappa}_q)|0\rangle\langle 0|+\tilde{\kappa}_q|1\rangle\langle 1|\right]$. Then, like in Ref.~\cite{HHL} again, we uncompute $|\tilde{\kappa}_q\rangle$ by undoing the phase estimation to get: $\frac{1}{2^N}\sum_q|\tilde{\xi}_q\rangle\langle\tilde{\xi}_q|\otimes\left[(1-\tilde{\kappa}_q)|0\rangle\langle 0|+\tilde{\kappa}_q|1\rangle\langle 1|\right]$. Finally, we measure the ancilla qubit and perform postselection, conditioned on the measurement outcome being $1$, to get the estimate $\tilde{\rho}$ of $\rho$. Since the probability of obtaining $1$ is $\sum_q\tilde{\kappa}_q=1$, the postselection is efficient. 

\section{Algorithm}
Our algorithm is as follows:
\begin{enumerate}
\item Construct the unitary $\mathcal{V}=e^{-i\rho\sqrt{\epsilon}}$ upto accuracy $\epsilon$ from the given state $\rho$, by applying (\ref{eq:exprho_step}) once (with $\nu\Delta t$ in place of $\Delta t$, where $\nu = \frac{\hbar\epsilon\sqrt{\epsilon}}{2 \pi^2}$), using a known diagonal matrix $\sigma$ with diagonal elements $\{\lambda_k\}$. \label{algo:step1}

\item Construct the unitary $\mathcal{U}=e^{-i\eta\sqrt{\epsilon}}$ with a single copy of $\eta=\mathcal{V}\sigma\mathcal{V}^{-1}$ similarly, where the eigenvalues of $\eta$ are $\{\lambda_k\}$. \label{algo:step2}

\item Apply to the initial state $|\Lambda_0\rangle|\lambda_k\rangle$, the operator $\mathcal{R}^\prime=\left(\mathcal{I}\otimes\mathcal{Q}\right)c\mathcal{U}^\prime\left(\mathcal{TG}^{-1}\otimes\mathcal{G}\mathcal{T}^{-1}\right)$ to get the state $|\tilde{\chi}_k\rangle|\lambda_k\rangle$ for each $k$. Here, we have $|\Lambda_0\rangle=\mathcal{G}|0\rangle=\sqrt{\frac{2}{T}}\sum_{\tau=0}^{T-1}\sin\frac{\pi(\tau+\frac{1}{2})}{T}|\tau\rangle$ for some large $T=2^N$, $\mathcal{T}:|0\rangle\langle 0|\to\mathcal{I}/2^N=\sum_l\mathcal{L}_l|0\rangle\langle 0|\mathcal{L}_l^\dagger$ with $\sum_l\mathcal{L}_l^\dagger\mathcal{L}_l=\mathcal{I}$, $c\mathcal{U}^\prime=\sum_{\tau=0}^{T-1}\mathcal{U}^{\tau/T} \otimes|\tau\rangle\langle\tau|$, $\mathcal{Q}$ is quantum Fourier transform with Fourier basis states $|p\rangle$, where $p=\frac{\pi\tilde{\lambda}_p}{\epsilon}$, and $\mathcal{I}$ is the identity operator. We measure the first register to get the estimates $|\tilde{\chi}_k\rangle$ of the eigenstates $|\chi_k\rangle$ of $\eta$, that form the columns of $\tilde{\mathcal{V}}$. \label{algo:step3}

\item Perform phase estimation of the obtained estimate $\tilde{\mathcal{V}}$ of unitary $\mathcal{V}$ with $\mathcal{I}/2^N=\frac{1}{2^N}\sum_q|\tilde{\xi}_q\rangle\langle\tilde{\xi}_q|$ as the input, to obtain at the output: $\frac{1}{2^N}\sum_q|\tilde{\kappa}_q\rangle\langle\tilde{\kappa}_q|\otimes|\tilde{\xi}_q\rangle\langle\tilde{\xi}_q|$, where $\tilde{\kappa}_q$ are the estimates of the eigenvalues $\kappa_q$ of $\rho$, and $|\tilde{\xi}_q\rangle$ are the eigenstates of $\tilde{\mathcal{V}}$ and are estimates of the eigenstates $|\xi_q\rangle$ of $\rho$ and $\mathcal{V}$. \label{algo:step4}

\item Add an ancilla qubit initialised in the maximally mixed state $\mathcal{I}/2=\frac{1}{2}(|0\rangle\langle 0|+|1\rangle\langle 1|)$, and rotate it, conditioned on $|\tilde{\kappa}_q\rangle$, to get: $\frac{1}{2^N}\sum_q|\tilde{\kappa}_q\rangle\langle\tilde{\kappa}_q|\otimes|\tilde{\xi}_q\rangle\langle\tilde{\xi}_q|\otimes\left[(1-\tilde{\kappa}_q)|0\rangle\langle 0|+\tilde{\kappa}_q|1\rangle\langle 1|\right]$. \label{algo:step5}

\item Uncompute $|\tilde{\kappa}_q\rangle$ by undoing the phase estimation to get: $\frac{1}{2^N}\sum_q|\tilde{\xi}_q\rangle\langle\tilde{\xi}_q|\otimes\left[(1-\tilde{\kappa}_q)|0\rangle\langle 0|+\tilde{\kappa}_q|1\rangle\langle 1|\right]$, and measure the ancilla qubit and perform postselection, conditioned on the measurement outcome being $1$, to get the estimate $\tilde{\rho}=\sum_q\tilde{\kappa}_q|\tilde{\xi}_q\rangle\langle\tilde{\xi}_q|$ of the state $\rho=\sum_q\kappa_q|\xi_q\rangle\langle\xi_q|$. \label{algo:step6}
\end{enumerate}

\section{Discussion}
\begin{figure}[!t]
    \centering
    \includegraphics[width=0.9\linewidth]{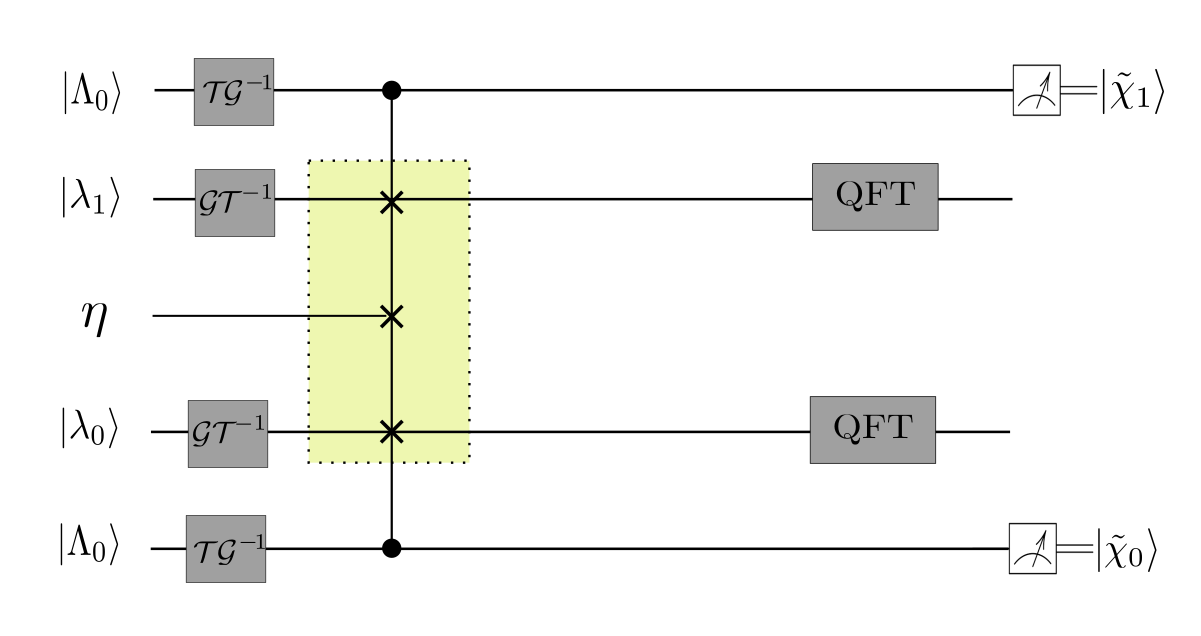}
    \caption{Parallel quantum phase estimations for all eigenstates of $\eta$ at once.}
    \label{fig:qpe3}
\end{figure}

Notice that in the algorithm from Ref.~\cite{BHMT} discussed earlier, it is required to prepare the second register to the state $|\psi\rangle$ every time we perform estimation of an amplitude. Because of no-cloning theorem, this implies that the entire quantum experiment is required to be repeated for estimation of every probability amplitude of the state. That is, if there are $\mathcal{N}$ probability amplitudes of the state to estimate, the algorithm, and therefore, the experiment, needs to be run $\mathcal{N}-1$ times (the last amplitude can be deduced from the fact that the squares of all $\mathcal{N}$ amplitudes sum to unity). In our approach, when we run phase estimation in step $3$, we initialize the register, other than the one used to store the phase estimate, to a known eigenstate of a unitary, obtained by exponentiation of a density matrix using just a single copy of the state. So, if there are $\mathcal{N}=2^N$ eigenstates of the unitary, one needs to perform the phase estimation process $\mathcal{N}$ times. However, it turns out that the phase estimations for all $k$ in step $3$ of our algorithm can be done in parallel. Consider that there are only two eigenvalues to estimate for eigenstates $|\lambda_0\rangle$ and $|\lambda_1\rangle$ for simplicity, that can be easily generalized to any number of eigenstates. We define $\mathcal{S}_0:=\mathcal{S}_{0\eta}\otimes\mathcal{I}_1$ and $\mathcal{S}_1:=\mathcal{I}_0\otimes \mathcal{S}_{\eta 1}$, and perform:
\begin{equation*}
    \left(e^{-i(\mathcal{S}_0+\mathcal{S}_1)\sqrt{\epsilon}}\right)\left(|\lambda_0\rangle\langle\lambda_0|\otimes\eta\otimes|\lambda_1\rangle\langle\lambda_1|\right)\left(e^{i(\mathcal{S}_0+\mathcal{S}_1)\sqrt{\epsilon}}\right),
\end{equation*}
as part of step $2$, where $\mathcal{S}_{0\eta}$ swaps $|\lambda_0\rangle$ and $\eta$, and $\mathcal{S}_{\eta 1}$ swaps $\eta$ and $|\lambda_1\rangle$. Also, $\mathcal{I}_0$ and $\mathcal{I}_1$ are identity operators acting on $|\lambda_0\rangle$ and $|\lambda_1\rangle$, respectively. Clearly, the operator $\mathcal{S}_0+\mathcal{S}_1$ is a sparse matrix, and therefore, its exponentiation can be simulated efficiently as before \cite{BACS,HHL}. We can then perform phase estimations of both the eigenstates at the same time in parallel as shown in Figure \ref{fig:qpe3}. Notice that an additional constant global phase factor arises for each eigenstate from the identity operators $\mathcal{I}$ within $\mathcal{S}_0$ and $\mathcal{S}_1$, that may be ignored or undone for in the final phase estimate for each eigenstate. Also, note from (\ref{eq:exprho_step}) that the original state $\rho$ itself is not destroyed, since we never perform measurements on register $C$. It undergoes a rotation to $\mathcal{E}\rho\mathcal{E}^\dagger$, where $\mathcal{E}:=e^{-i\sigma\sqrt{\epsilon}}$. Since we know the state $\sigma$, we can keep $\rho$ unchanged for any further quantum processing by applying $\mathcal{E}^\dagger\mathcal{E}\rho\mathcal{E}^\dagger\mathcal{E}=\rho$. Similarly, since we know $|\lambda_k\rangle$ for each $k$ and $|\Lambda_0\rangle$ in Figure \ref{fig:qpe3}, the same copy of the state $\eta$ can be recycled to create the controlled unitary $c\mathcal{U}^\prime$ with varying times, as desired for each phase estimation in step $3$, by undoing the unitary evolution $e^{-i\sum_k|\lambda_k\rangle\langle\lambda_k|\sqrt{\epsilon}}$ (which is efficient to simulate, since $\sum_k|\lambda_k\rangle\langle\lambda_k|$ is diagonal, and so sparse, with orthonormal eigenvectors $\{|\lambda_k\rangle\}$) that $\eta$ undergoes conditioned on $|\Lambda_0\rangle$.

Thus, we perform quantum state tomography: (i) using a single copy of the unknown arbitrary state, (ii) without knowing the Hamiltonian, i.e.~evolution dynamics, of the state, (iii) without destroying the state with measurements, that can be recovered for any further use. To the best of our knowledge, no existing algorithm performs quantum state tomography, ensuring all three. For example, Ref.~\cite{FGHLNS} requires the Hamiltonian to be known to perform single-copy state tomography. Ref.~\cite{OMES} requires a prior information that the state is pure, or close to pure. Moreover, like in compressed sensing, they require the density matrix to be sparse. Our algorithm works for any density matrix, sparse or not, pure or noisy, low rank or not. Ref.~\cite{MSC} requires a pointer state (a kind of ancilla like in Ref.~\cite{OMES}), and performs two-state tomography on the combined system. We do not measure the original state, that just undergoes a rotation, and can be restored easily. As an example, a typical quantum image processing experiment \cite{YWLCP} is required to be repeated several times, to extract the information from $\mathcal{N}$ pixels of the output image, using existing quantum amplitude estimation or quantum state tomography techniques \cite{VAB,RCLT}. By contrast, our approach allows for the information from $\mathcal{N}$ pixels of the output image of the experiment to be inferred from only one run of the experiment, followed by an approximate estimation of the image state's density matrix.

Clearly, our algorithm works only for non-zero error $\epsilon$ in the simulations of the density matrix exponentiations. The error can be very small, but never zero, otherwise our algorithm would not work. This is unavoidable, since if our algorithm worked for zero error, it would violate the no-cloning theorem of quantum mechanics \cite{DJG}. Quantum no-cloning theorem forbids only perfect cloning (i.e.~with zero error) of arbitrary quantum states. The fact that the algorithm cannot work for zero error is precisely why the algorithm does not violate the no-cloning theorem. This is similar to imperfect quantum state cloning in the existing literature, such as Ref.~\cite{PWZLLH}. As long as the error is non-zero, our method does not violate/override any fundamental theorem.

It is possible that our algorithm can be used to break the quantum key distribution algorithm, since the original state is not destroyed by any measurement. However, notice that if we use too small an error to be able to reconstruct the original state as closely as possible, our algorithm will not be efficient anymore. Our algorithm requires that we use an $\epsilon$ that is $O(1/{\rm poly}N)$ for our algorithm to be efficient, but one might need an $\epsilon$ that is exponentially small in $N$, i.e.~$O(1/2^N)$, in order for our algorithm to break the quantum key distribution algorithm, unless it is enough to reconstruct the original state only approximately. 
A precision error of $O(1/2^N)$ is ideally required to be able to distinctly resolve all the $2^N$ eigenvalues of $\rho$, either when simulating the exponentiation of the density matrix $\rho$, or when performing phase estimation on the resulting unitary $\mathcal{V}:=e^{i\rho t}$ to capture the eigenvalues as phase estimates, which will then require exponential time. 
This is why we relax the requirement of having to resolve all the eigenvalues distinctly, and take the precision to be $\epsilon=O(1/{\rm poly}N)$. 
This is acceptable for us, because in the improved quantum phase estimation that we use, the quantity $\epsilon/2$ gives the precision error in trace distance, which is the maximum probability of estimation error, that need not be exponentially small, as long as it is as small as is required for the total accumulated probability of error of our overall algorithm to be less than or equal to $1/3$.

We are doing quantum state tomography and not really cloning here. If we were to do cloning here, we could directly go to step \ref{algo:step4} after step \ref{algo:step1} of our algorithm. That is, once we have exponentiated $\rho$ to get the unitary $\mathcal{V}$, we can directly perform phase estimation on it (instead of $\tilde{\mathcal{V}}$) as in step \ref{algo:step4}, then perform step \ref{algo:step5}, and then get a copy of $\rho$ reconstructed in step \ref{algo:step6}, without having destroyed the original copy of $\rho$. Of course, the copy would not be exact replica of the original state, because otherwise it would violate the no-cloning theorem. The copy can only approximately resemble the original state because 
the precision cannot be exponentially small in $N$ and can only be $O(1/{\rm poly}N)$, since otherwise the process would take exponential time. The steps \ref{algo:step2} and \ref{algo:step3} of our algorithm enable us to find an estimate $\tilde{\mathcal{V}}$ of $\mathcal{V}$, before we go on to reconstruct an estimate of $\rho$ from $\tilde{\mathcal{V}}$. Thus, a caveat is that our method works only for non-zero error, polynomially, and not exponentially, small in $N$, that is acceptable for us.

Note that in Figure \ref{fig:qst}, we show how the state $\zeta$ can be prepared in two ways from the initial state $|00\rangle$. The top flow is where we know the eigenstates $\chi$ and we estimate the eigenvalues $\lambda$ (the usual thing done in phase estimation), and the bottom flow is where we know the eigenvalues $\lambda$ and we estimate the eigenstates $\chi$. In our algorithm, we know the eigenvalues of $\eta$ in our step \ref{algo:step2}, but we need to find the eigenstates $\chi$ in our step \ref{algo:step3}. Since in Figure \ref{fig:qst}, the top and bottom flows can be treated as equivalent, we achieve the bottom flow (the task of estimating the eigenstates) by using the top flow in a certain manner. That is, the controlled-$\mathcal{W}$ operations required for the bottom flow are realized in terms of the controlled-$\mathcal{U}$ operations from the top flow. Of course, we need to know the individual eigenstates, which we cannot know from all of them in a superposition as in $\zeta$. This is why, we perform step \ref{algo:step3} of our algorithm as in our Figure \ref{fig:qpe3}, so that we can estimate each of the $2^N$ number of eigenstates individually, but all in quantum-parallel. This is made possible by means of sums of swap operators, when exponentiating $\eta$ in step \ref{algo:step2}, as outlined earlier. Also, notice that Ref.~\cite{KLLOY} discusses optimal bounds on errors or copies needed, as applicable to the density matrix exponentiation (DME) task, that was originally proposed by LMR in Ref.~\cite{LMR}. Our work does not violate any of these relevant bounds. For example, the optimal number of copies of the state required for DME is $O(t^2/\epsilon)$, where if we simply take $t=O(\sqrt{\epsilon})$, we would have $O(t^2/\epsilon)=O(1)$. While this implies that only a single copy is required, it clearly does not violate the optimal sample complexity for the task. So, nothing fundamental is broken in our work.

\begin{figure}[!t]
    \centering
    \includegraphics[width=0.9\linewidth]{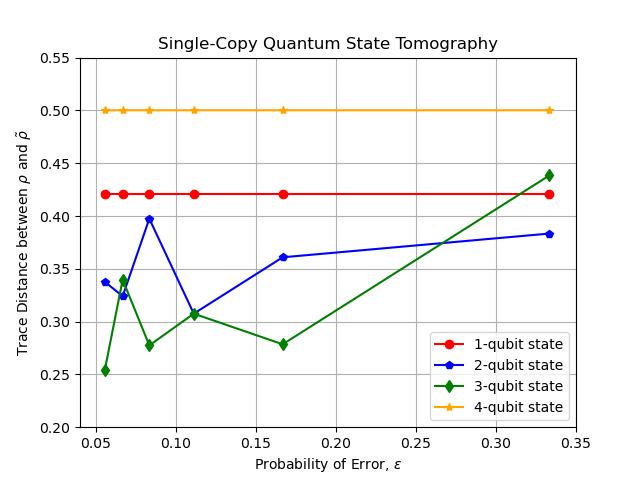}
    \caption{Trace distance between $\rho$ \& $\tilde{\rho}$ vs.~$\epsilon$ for $1$, $2$, $3$ and $4$ qubit states.}
    \label{fig:eqst}
\end{figure}

We simulated our algorithm to estimate high rank and dense density matrices of $1$, $2$, $3$, and $4$ qubit states with eigenvalues $\{0.723606$, $0.276393\}$, $\{0.295570$, $0.257141$, $0.211747$, $0.235541\}$, $\{0.12$, $0.114644$, $0.139193$, $0.151533$, $0.144628$, $0.11$, $0.115$, $0.105\}$ and $\{0$, $0$, $0$, $0.0172$,  $0.023$, $0.0287$, $0.0345$, $0.046$, $0.0575$, $0.069$, $0.0805$, $0.092$, $0.1034$, $0.1264$, $0.1494$, $0.1724\}$, respectively. Figure \ref{fig:eqst} shows the plot of the trace distance between each of these states and its estimate, calculated for $\epsilon$ equal to $1/3$, $1/6$, $1/9$, $1/12$, $1/15$ and $1/18$. Clearly, each state has the lowest trace distance with its estimate for at least the value of $\epsilon$, the twice of which is closest to the smallest non-zero eigenvalue of that state.

\section{Conclusion}
We presented an efficient quantum state tomography algorithm to estimate a quantum state with a single copy of the state, without knowing its evolution dynamics and without destroying it. Note that even if we use a non-zero value of $\epsilon$, it may appear that our algorithm can still perfectly clone the given unknown arbitrary quantum state, if one is allowed to repeat our algorithm an \emph{infinite} number of times. However, this is not possible because of abstraction of infinite precision even in classical physics, as Ref.~\cite{FDSNG} argues that no-cloning also applies to statistical ensembles of classical systems.




\bibliographystyle{IEEEtran}
\bibliography{qstbib}

\end{document}